\documentclass[aps,pra,twocolumn,amsmath,amssymb,showpacs]{revtex4-1}
\usepackage{graphicx}
\usepackage{dcolumn}
\usepackage{bm}
\usepackage{hyperref}
\usepackage[mathlines]{lineno}
\usepackage{color}

\begin{document}


\title{Asymmetric sequential Landau-Zener dynamics of Bose condensed atoms in a cavity}

\author{Jiahao Huang$^{1,2}$}
\author{Pu Gong$^{3}$}
\author{Xizhou Qin$^{1}$}
\author{Honghua Zhong$^{1,4}$}

\author{Chaohong Lee$^{1,2,}$}%
\altaffiliation{Emails: lichaoh2@mail.sysu.edu.cn; chleecn@gmail.com}

\affiliation{$^{1}$School of Physics and Astronomy, Sun Yat-Sen University, Zhuhai 519082, China}

\affiliation{$^{2}$State Key Laboratory of Optoelectronic Materials and Technologies, Sun Yat-Sen University, Guangzhou 510275, China}

\affiliation{$^{3}$Department of Physics, and Center for Theoretical and Computational Physics, The University of Hong Kong, China}

\affiliation{$^{4}$Department of Physics, Jishou University, Jishou 416000, China}

\date{\today}

\begin{abstract}
  We explore the asymmetric sequential Landau-Zener (LZ) dynamics in an ensemble of interacting Bose condensed two-level atoms coupled with a cavity field.
  Assuming the couplings between all atoms and the cavity field are identical, the interplay between atom-atom interaction and detuning may lead to a series of LZ transitions.
  Unlike the conventional sequential LZ transitions, which are symmetric to the zero detuning, the LZ transitions of Bose condensed atoms in a cavity field are asymmetric and sensitively depend on the photon number distribution of the cavity.
  In LZ processes involving single excitation numbers, both the variance of the relative atom number and the step slope of the sequential population ladder are asymmetric, and the asymmetry become more significant for smaller excitation numbers.
  Furthermore, in LZ processes involving multiple excitation numbers, there may appear asymmetric population ladders with decreasing step heights.
  During a dynamical LZ process, due to the atom-cavity coupling, the cavity field shows dynamical collapse and revivals.
  In comparison with the symmetric LZ transitions in a classical field, the asymmetric LZ transitions in a cavity field originate from the photon-number-dependent Rabi frequency.
  The asymmetric sequential LZ dynamics of Bose condensed atoms in a cavity field may open up a new way to explore the fundamental many-body physics in coupled atom-photon systems.
\end{abstract}


\maketitle

\section{Introduction}\label{Sec1}

The great achievements in manipulating and probing ultracold atoms in a cavity offer a new platform for exploring many-body systems and their dynamics.
Up to now, the strong atom-cavity coupling has been experimentally demonstrated in several laboratories by using an atomic Bose-Einstein condensate (BEC)~\cite{Brennecke2007, Colombe2007, Goldwin2011, Baumann2011, Ritsch2013, Chen2015}.
The BEC-cavity experiments can be used to study the coupled atom-photon system with both large and small atom and photon numbers~\cite{Colombe2007, Brennecke2007, Goldwin2011, Chen2015}.
The BEC-cavity system provides an ideal platform for exploring exotic many-body quantum effects, such as quantum phase transition~\cite{Mekhov2007, Ritsch2013,Larson2008, Nagy2010, Baumann2010, Baumann2011, Vacanti2012, Bakhtiari2015}, many-body quantum entanglement~\cite{Haas2014, Sun2014, McConnell2015}, precision measurement~\cite{Zuppardo2015, Hamilton2015}, and many-body quantum dynamics~\cite{Alvermann2012, Garraway2012}.
On the other hand, as an important fundamental problem, the many-body Landau-Zener (LZ) tunneling with BEC had attracted much attention for a long time~\cite{Wu2000,Chen2011, Qian2013, Zhong2014}.
However, the many-body LZ tunneling in BEC-cavity system, in particular, how the cavity field affects the LZ dynamics of atoms, is still unclear.

Due to the atom-atom interaction, the many-body LZ tunneling is very different from the single-particle one.
The sequential LZ tunneling induced by interaction blockade have been found in Bose-Josephson system (BJS)~\cite{Lee2008, Cheinet2008}, Bose-Hubbard ladder~\cite{Deng2015} and spinor BEC within optical superlattices~\cite{Wagner2011}.
For an example, in a BJS~\cite{Gati2007, Lee2009, Gross2010, Lee2012}, due to the interplay between inter-mode bias and atom-atom interaction, the resonant tunneling and interaction blockade take place~\cite{Lee2008, Cheinet2008}.
There appears a population ladder indicating a series of interaction blockade.
The variance of the relative population would exhibit several peaks representing the single-atom resonant tunneling.
Due to the fixed Josephson coupling in such a BJS, the step slopes of the ladder and the resonant tunneling peaks are symmetric about the zero bias (which corresponds to the zero detuning).

In a BEC-cavity system, due to the quantization nature of the cavity field, the picture of the sequential LZ dynamics may have significant differences.
In comparison with the BJS, the Rabi frequency in a BEC-cavity system acts the role of the Josephson coupling.
Unlike the Josephson coupling, the Rabi frequency is no longer fixed and it is proportional to the square root of the cavity-photon number~\cite{Fink2008}.
In the case of few photons, as the photon number changes in the LZ process, the Rabi frequency dramatically changes when the detuning is tuned.
Furthermore, if there is no sufficient photons to excite all the atoms to the upper level, an incomplete sequential LZ process would occur.
Thus, in addition to the atom-atom interaction and the detuning, the photon number also plays an important role in the LZ process of the BEC-cavity system.
Simultaneously, during the time-evolution, the atoms may also have an influence on the cavity field.
Therefore, it is of great interest to investigate: (i) how the sequential LZ dynamics is affected by the cavity field? and (ii) how the cavity field changes during such a LZ process?

In this article, we study the LZ process of an ensemble of interacting Bose condensed atoms trapped in a cavity.
In the framework of second quantization, the BEC-cavity system can be described by a modified Tavis-Cummings model with an additional nonlinear term determined by atom-atom interaction~\cite{Garraway2012, Feng2015}.
If the initial cavity field is described by a Fock state, the BEC-cavity system only involves a single excitation number.
In the adiabatic sweeping of the detuning, the interplay between the detuning and the atom-atom interaction results in the asymmetric sequential LZ transitions, in which the atoms absorb or emit photons one by one and the sequential LZ transitions are asymmetric.
If the initial cavity field is described by a coherent state, the BEC-cavity system will involve multiple excitation numbers.
Different from the case of single excitation numbers, in the asymmetric sequential LZ process, the atoms may no longer absorb entirely one photon for every population step.
The amount of absorbed/emmited photons becomes less than one and it gradually decreases even in the adiabatic limit.
During the LZ process, the cavity field is no longer a coherent state and its intermittent collapse and revival appear.

This article is organized as follows.
In Sec.~\ref{Sec2}, we describe the model and give its Hamiltonian.
In Sec.~\ref{Sec3}, we investigate the asymmetric sequential LZ process involving single excitation numbers.
In Sec.~\ref{Sec4}, we consider the atoms initially coupled with a coherent cavity field, and then study the asymmetric sequential LZ process involving multiple excitation numbers. We derive an analytical formula for sequential ladders of the photon number and the relative atom number. The adiabaticity conditions for the sequential LZ transitions are also analytically estimated. An intriguing findings of fractional steps are observed when the average number is small compared with atom number. In addition, we analyze the effects of the cavity photon loss on the sequential LZ dynamics. The phenomena of collapse and revivals of the cavity field are also revealed.
Finally, in Sec.~\ref{Sec5}, we give a brief summary and discussion.

\section{Model}\label{Sec2}

\begin{figure}[htb]
\centering
\includegraphics[width=\columnwidth]{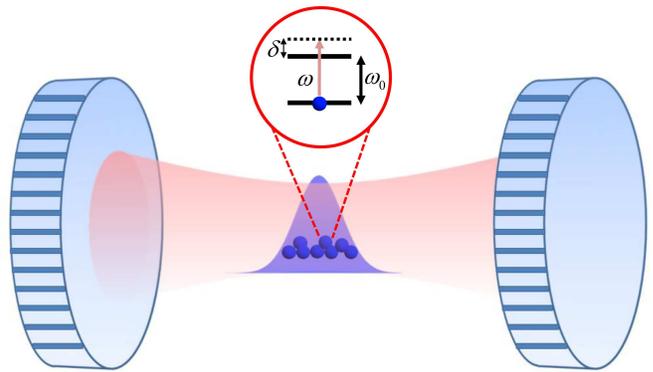}\caption{(Color online) Sketch of interacting Bose condensed atoms trapped in an optical cavity. Each atom is identically coupled to the cavity mode with coupling strength $g/\sqrt{N_a}$. Here, the frequencies of the cavity field and the atomic transition are denoted by $\omega$ and $\omega_0$, respectively. The detuning is $\delta=\omega-\omega_0$ and the effective atom-atom interaction is characterized by $E_c$.}
\label{Fig-schematic}
\end{figure}

We consider an ensemble of Bose condensed atoms occupying two different hyperfine levels and trapped in an optical cavity~\cite{Colombe2007, Brennecke2007}.
The sketch is shown in Fig.~\ref{Fig-schematic}.
The atoms are coupled with a cavity mode, in which one atom in the lower level may absorb a photon and jump to the upper level, or may be taken from the upper level into the lower level with a photon emitted.
Assuming that the coupling fields are spatially uniform and the atom-atom collision do not change the internal states. Each atom identically couples with a single-mode cavity cavity field~\cite{Lee2014}. Therefore, the system obeys a second quantized Hamiltonian within a single-mode approximation (in the unit of $\hbar=1$ throughout the paper)
\begin{eqnarray}\label{H1}
    H &=& \omega \hat a^{\dagger} \hat a +\sum_{j=\uparrow,\downarrow}(E_{0j} \hat b^{\dagger}_{j} \hat b_{j} + \frac{1}{2}G_{jj} \hat b^{\dagger}_{j} \hat b^{\dagger}_{j} \hat b_{j} \hat b_{j})\nonumber \\
    &+& G_{\uparrow \downarrow}\hat b^{\dagger}_{\downarrow} \hat b^{\dagger}_{\uparrow} \hat b_{\uparrow} \hat b_{\downarrow} + \frac{g}{\sqrt{N_a}}(\hat a \hat b^{\dagger}_{\uparrow} \hat b_{\downarrow}+\hat a^{\dagger} \hat b^{\dagger}_{\downarrow} \hat b_{\uparrow}),\\
    &&\nonumber
\end{eqnarray}
where $\omega$ is the frequency of cavity mode, $N_a=N_{\uparrow}+N_{\downarrow}=\hat b^{\dagger}_{\uparrow} \hat b_{\uparrow} + \hat b^{\dagger}_{\downarrow}\hat b_{\downarrow}$ is the atom number,  $E_{0j}$ is the hyperfine energy of the atoms in state $|j\rangle$, $G_{jj}$ is the atom-atom interaction energies of state $|j\rangle$, $G_{\uparrow\downarrow}$ is the atom-atom interaction energy between states $|\uparrow \rangle$ and $|\downarrow \rangle$, and $g$ is the homogenous coupling strength between the cavity mode and the Bose condensed atoms.

Regarding all the atoms as spin-1/2 particles, one can define the angular momentum operators as
\begin{eqnarray}\label{JxJyJz}
  \hat J_{x} &=& \frac{\hat b^{\dagger}_{\uparrow} \hat b_{\downarrow}+ \hat b_{\uparrow} \hat b^{\dagger}_{\downarrow}}{2}, \\
  \hat J_{y} &=& \frac{\hat b^{\dagger}_{\uparrow} \hat b_{\downarrow}- \hat b_{\uparrow} \hat b^{\dagger}_{\downarrow}}{2i}, \\
  \hat J_{z} &=& \frac{\hat b^{\dagger}_{\uparrow} \hat b_{\uparrow}- \hat b^{\dagger}_{\downarrow} \hat b_{\downarrow}}{2}.
\end{eqnarray}

The system has two good quantum numbers: the atom number $N_a$ and the  excitation number $N_e=\hat a^{\dagger} \hat a+N_{\uparrow}=\hat a^{\dagger} \hat a+\hat J_z+\frac{N_a}{2}$, satisfying the commutation $[N_{a},H]=0$ and $[N_{e},H]=0$.
Here, when the excitation number is much larger than the atom number, i.e., $N_e \gg N_a$, it corresponds to the large excitation number condition, which is close to the classical field limit.
While if the excitation number is comparable with or smaller than the atom number, i.e., $N_e \lesssim N_a$, it refers to the small excitation number condition.
The quantization of cavity mode would make a difference with a classical field and the photon number may affect the properties of the whole system.
We will carefully compare the results under large and small excitation conditions in the following.

Since $N_a$ and $N_e$ are conserved quantities, the constant terms $\mathcal{O}(N_a)$ and $\mathcal{O}(N^2_a)$ can be eliminated and the Hamiltonian becomes
\begin{equation}\label{H2}
    H = \omega_0 \hat J_z + \omega \hat a^{\dagger} \hat a + \frac{E_c}{2}\hat J^2_{z} +\frac{g}{\sqrt{N_a}}(\hat a^{\dagger} \hat J_{-}+\hat a \hat J_{+}),
\end{equation}
where $\hat J_{\pm}=\hat J_x \pm i\hat J_y$ are the raising and lowering operators of atoms, the atomic transition frequency $\omega_0=E_{0\uparrow}-E_{0\downarrow}+\frac{1}{2}(N_a-1)(G_{\uparrow\uparrow}-G_{\downarrow\downarrow})$, and the effective atom-atom interaction $E_c = G_{\uparrow\uparrow}+G_{\downarrow\downarrow}-2G_{\uparrow\downarrow}$.
Further, shifting the zero point of the Hamiltonian~\eqref{H2}, one can get
\begin{equation}\label{H3}
    H = -\frac{\delta}{2} \hat J_z + \frac{\delta}{2} \hat a^{\dagger} \hat a + \frac{E_c}{2}\hat J^2_{z} +\frac{g}{\sqrt{N_a}}(\hat a^{\dagger} \hat J_{-}+\hat a \hat J_{+}),
\end{equation}
with the detuning $\delta=\omega-\omega_0$. According to the last term, for fixed $g$ and $N_a$, the Rabi frequency is proportional to the square root of photon number (i.e., $\Omega\propto\sqrt{n}$), which is changeable with respect to the state of cavity field.

The state of the system can be expanded as
\begin{equation}\label{Psi}
    |\psi\rangle=\sum_{n,m} C_{n,m} |n,m\rangle,
\end{equation}
where $n$ is the photon number of the Fock state, $m=(N_{\uparrow}-N_{\downarrow})/2$ is the relative atom number and $C_{n,m}$ are the coefficients of basis $|n,m\rangle$.
Due to the conservation of excitation number, when $N_e < N_a$, the number of bases can be reduced to $N_e+1$, the state of the system can be expanded as
\begin{equation}\label{Psi1}
    |\psi\rangle=\sum_{l=0}^{N_e} C_{l}^{N_e} |N_e-l,-\frac{N_a}{2}+l\rangle.
\end{equation}
While for $N_e \ge N_a$, only $N_a+1$ bases are enough, and the state of the system can be expanded as
\begin{equation}\label{Psi2}
    |\psi\rangle=\sum_{l=0}^{N_a} C_{l}^{N_e} |N_e-l,-\frac{N_a}{2}+l\rangle.
\end{equation}
Here, $C_l^{N_e}$ in Eq.~\eqref{Psi1} and~\eqref{Psi2} denotes the coefficient with $N_e-l$ photons and $l$ atoms populating in the upper level.

\section{Landau-Zener Process involving single excitation numbers}\label{Sec3}

Firstly, we consider the system with fixed single excitation numbers and study the static properties of the system in small and large excitation number conditions.
Starting from Hamiltonian~\eqref{H3}, we choose atom number $N_a=10$ and investigate the situations with excitation numbers $N_e=5$, $N_e=10$ and $N_e=1000$. Here, the system state $|\Psi\rangle$ can be expressed according to Eq.~\eqref{Psi1} or Eq.~\eqref{Psi2}. We vary $\delta$ from negative detuning to positive detuning and obtain the corresponding groundstates.
The detuning $\delta$ tends to populate the atoms into the upper level while the repulsive atom-atom interaction ($E_c>0$) tends to populate the atoms equally in both levels.
For far negative detuning, i.e., $\omega \ll \omega_0$, $\delta\rightarrow -\infty$, the groundstate is $|N_e,-\frac{N_a}{2}\rangle$, which corresponds to that all atoms are in the lower level and the photon number equals to the excitation number $N_e$.
For far positive detuning, i.e., $\omega \gg \omega_0$, $\delta\rightarrow \infty$, the form of groundstate is related to the excitation number. When $N_e\ge N_a$, the groundstate is $|N_e-N_a,\frac{N_a}{2}\rangle$, which is the state that all atoms occupy in the upper level and the photon number becomes $N_e-N_a$. When $N_e<N_a$, it becomes $|0,-\frac{N_a}{2}+N_e\rangle$, where only a portion of atoms occupy in the upper level and no photons remain.
In other regions of detuning, the groundstate depends sensitively on the parameters $g$, $\delta$ and $E_c$.
For better revealing the properties of the system, the expectations of photon number $\langle n \rangle=\langle \psi|\hat a^{\dagger} \hat a |\psi\rangle$ and the half relative atom number $\langle N_{\uparrow}-N_{\downarrow}\rangle/2=\langle \psi| \hat J_z |\psi\rangle$, and the variance of relative atom number $\text{Var} (N_{\uparrow}-N_{\downarrow})=4(\langle \hat J_{z}^2 \rangle-\langle \hat J_{z} \rangle^2)$ for the groundstate are calculated and shown in Fig.~\ref{Fig-single-excitation}.

\subsection{Asymmetric sequential population ladders}

\begin{figure*}[htb]
\centering
\includegraphics[scale=0.73]{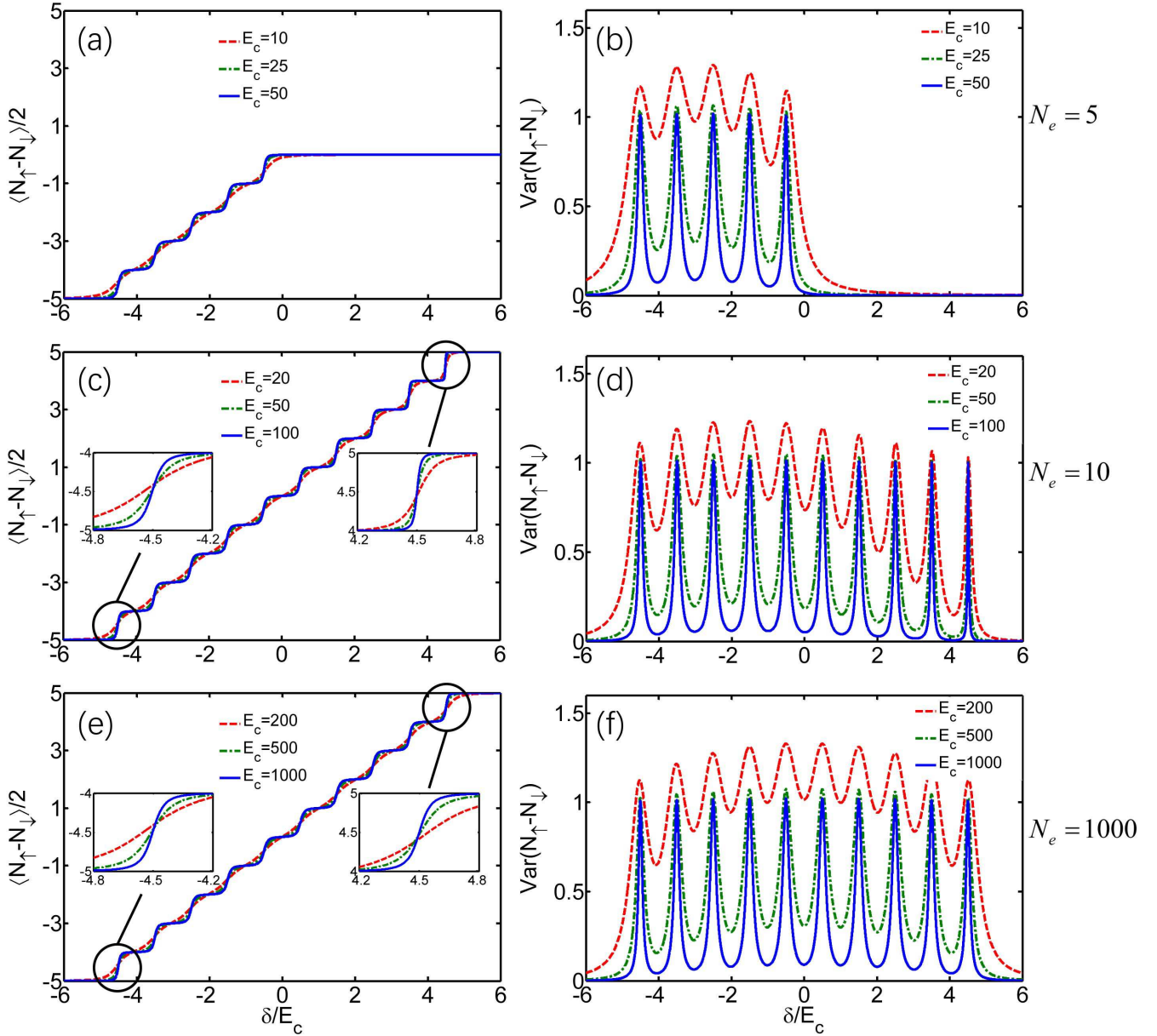}\caption{(Color online) Static properties of the system involving single excitation numbers. The top, middle and bottom rows correspond to the cases of $N_e=5$, $N_e=10$ and $N_e=1000$. The left and right columns represent the expectations of half relative atom number $\langle N_{\uparrow}-N_{\downarrow}\rangle/2=\langle \hat J_z \rangle$, and variance of relative atom number $\text{Var} (N_{\uparrow}-N_{\downarrow})=4(\langle \hat J_{z}^2 \rangle-\langle \hat J_{z} \rangle^2)$ versus detuning, respectively. Here, $N_a=10$ and $g=1$.}
\label{Fig-single-excitation}
\end{figure*}

There appear a series of ladder-like steps for the half relative atom number.
These sequential population ladders are induced by the atom-atom interaction. The steps become steeper and the plateaus get more smooth when the atom-atom interaction becomes larger.
The height of every step equals exactly 1. For every step, one of the atoms in the lower level can absorb one photon and be excited to the upper level.
The appearance of sequential population ladders is similar to the interaction blockade observed in the Bose-Josephson junctions~\cite{Lee2008, Cheinet2008}.
For a two-component Bose-Einstein condensate linked by classical Raman fields, the competition between detuning and nonlinear interaction will result in a sequential population ladder for the relative atom number, where the slopes of the steps are exactly symmetric about $\delta=0$.

However, for the two-state Bose condensed atoms coupled with a quantized cavity mode, the number of the photons make the shape of the ladders significantly different.
When the excitation number is smaller than atom number $N_e<N_a$, the structure is totally asymmetric since there are not enough photons to excite all the atoms and the population ladder ceases to increase when there are no photons inside the cavity, see Fig.~\ref{Fig-single-excitation}~(a).
In the situation of $N_e=10$, the step slopes of the population ladders are also no longer symmetric about $\delta=0$.
For the same magnitude of detuning $|\delta|$, the step slope under negative detuning $-\delta$ is always less sharper than the one under positive detuning $+\delta$.

The asymmetry of the step slopes comes from the changing Rabi frequency during the sequential LZ processes. Adiabatically sweeping the detuning from negative to positive, the photon number decreases downstairs from $N_e$ to 0. Therefore, the Rabi frequency $\Omega$ is getting smaller. The sharpness is determined by the ratio between Rabi frequency and the atom-atom interaction $\Omega/E_c$, where the smaller $\Omega/E_c$ corresponds to the sharper step slope. Specially, when the photon number decreases to zero, the Rabi frequency vanishes, and no transitions would happen.
Since the photon number under negative detuning $-\delta$ is always larger than the one under positive detuning $+\delta$, the step slopes of the ladder for $\langle N_{\uparrow}-N_{\downarrow}\rangle/2$ is asymmetric about $\delta=0$. The first and the last step slopes have the biggest difference because they respectively correspond to the largest and smallest photon number, see the insets of Fig.~\ref{Fig-single-excitation}~(c). While for other pairs of step slopes, the difference is smaller.

As the excitation number increases, the photon number difference under $-\delta$ and $+\delta$ becomes less dramatic and the Rabi frequency $\Omega$ tends to be uniform.
For large excitation number $N_e=1000$, the step slopes of the ladder become almost symmetric about $\delta=0$, see the insets of Fig.~\ref{Fig-single-excitation}~(e). This is because when the photon number is extremely large compared with the atom number, the cavity field can be treated as a classical field with fixed complex number, and the sequential behaviors would return to the symmetric one in a BJS.

\subsection{Asymmetric variance of relative atom number}

For further investigation, we look in the variance of relative atom number $\text{Var} (N_{\uparrow}-N_{\downarrow})$.
There are a series of peaks for $\text{Var} (N_{\uparrow}-N_{\downarrow})$. The locations of the peaks correspond to the locations of the step slopes in the ladder of $\langle N_{\uparrow}-N_{\downarrow}\rangle/2$. The stronger atom-atom interaction $E_c$ induces sharper and narrower peaks, which is consistent with the sharpness of the step slopes.
When $N_e=5$, the peaks are completely asymmetric. There appear several peaks when $\delta<0$ and no peaks for $\delta>0$, see Fig.~\ref{Fig-single-excitation}~(b).

The asymmetry is also obvious for $N_e=10$, see Fig.~\ref{Fig-single-excitation}~(d). The total structure is asymmetric about $\delta=0$. The width and height of first and the last peaks are totally different. The last peak is much sharper than the first one, which is in accordance with the situation of the step slopes of population ladder.
The plateaus in the ladder of $\langle N_{\uparrow}-N_{\downarrow}\rangle/2$ indicate the interaction blockades. In the regions of the plateaus, the system is approximately in the state $|n,m\rangle$ with definite photon number and relative atom number. Therefore, the variance of the relative atom number are suppressed.
The resonance peaks are caused by the quasi-degeneracy between the two states $|N_e-k+1, -\frac{N_a}{2}+k-1\rangle$ and $|N_e-k, -\frac{N_a}{2}+k\rangle$ with $k=1,...,\text{min}(N_a,N_e)$ in the neighboring plateaus. In the intermediate region of the step slopes, the superposition of these two states leads to $\text{Var} (N_{\uparrow}-N_{\downarrow})$ increasing dramatically. One can treat the atom-photon coupling term as a perturbation~\cite{Lee2008, Cheinet2008} and obtain the locations of the peaks with $\delta/E_c= l+1/2$ for $l=\{-N_a/2, -N_a/2+1, ..., N_a/2-1\}$, which is in agreement with the numerical results.

More importantly, since the Rabi frequency becomes weaker when the photon number decreases, the relative atom number also shows asymmetric properties. When the Rabi frequency decreases, the perturbation term becomes smaller, and the two neighboring states $|N_e-k+1, -\frac{N_a}{2}+k-1\rangle$ and $|N_e-k, -\frac{N_a}{2}+k\rangle$ are closer to degeneracy. Therefore, the variance of the relative atom number would appear narrower and sharper peaks at the degeneracy location.
While for large excitation number, the asymmetry becomes much smaller. Under $N_e=1000$, the structure of the variance $\text{Var} (N_{\uparrow}-N_{\downarrow})$ become more symmetric about $\delta=0$, see Fig.~\ref{Fig-single-excitation}~(f).

The set of peaks that indicate the single-atom resonance transition can be used to design the single-atom device. For this system, varying the photon number in the cavity can adjust the sensitivity of the response and the atom-atom interaction need not be too strong, which can be promising for application in sensitive detection and high-precision metrology.

\section{Landau-Zener dynamics involving multiple excitation numbers}\label{Sec4}

\begin{figure*}[htb]
\centering
\includegraphics[scale=0.6]{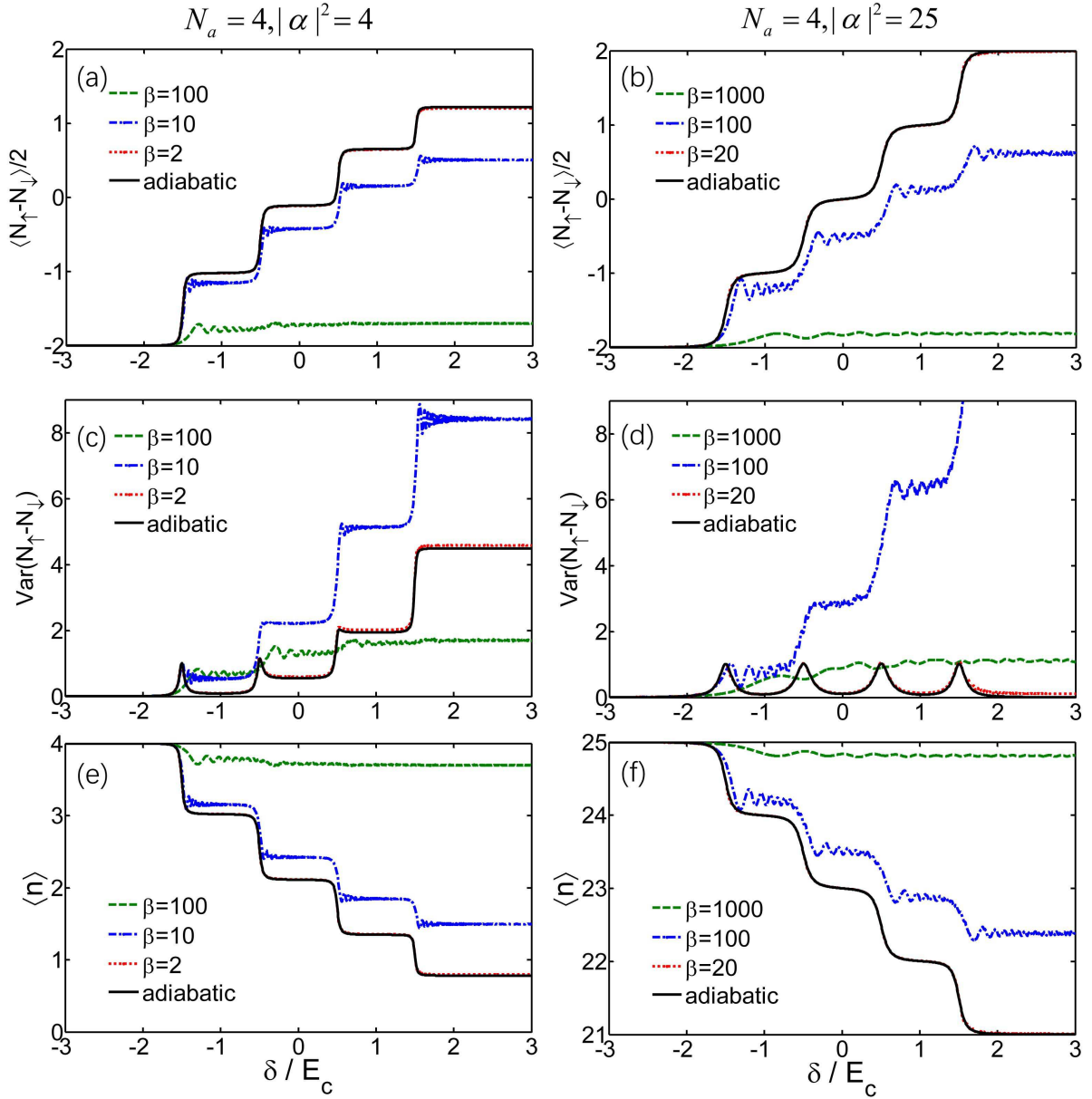}\caption{(Color online) Dynamical evolution of the half relative atom number, variance of relative atom number and photon number. The atom number $N_a=4$ and the initial cavity field is in a coherent state with average photon number $|\alpha|^2=4$ and $|\alpha|^2=25$. The detuning is linearly swept according to $\delta(t)=\delta_0+\beta t$. The black solid lines are the adiabatic limit and other lines are numerical results with different sweeping rates $\beta$. In order to ensure the initial detuning is large enough, we choose $\delta_0=-3 E_c$ with $E_c=100$ for calculation.
}
\label{Fig-Dynamical-LZ}
\end{figure*}

In the following, we consider the ensemble of interacting Bose condensed atoms coupled with a cavity field in a coherent state. The coherent state of photons is natural and it is easy to be prepared in experiments.
We sweep the detuning to investigate the LZ dynamics of the system.
Initially, the atoms are all prepared in the lower level. The cavity field is a coherent state, which is the eigenstate of annihilation operator $\hat a$ and can be written as $|\alpha\rangle = e^{-\frac{|\alpha|^2}{2}}\sum_{n=0}^{\infty} \frac{\alpha^n}{\sqrt{n!}}|n\rangle$,
with $|\alpha|^2$ being the average photon number of the coherent state.
Therefore, the initial state of the whole system can be expressed as
\begin{equation}\label{Psi0}
    |\Psi(0)\rangle=|\alpha, -\frac{N_a}{2}\rangle=\sum_{N_e=0}^{\infty} w_{N_e}|N_e,-\frac{N_a}{2}\rangle,
\end{equation}
which involves multiple excitation numbers with $w_{N_e}=e^{-\frac{|\alpha|^2}{2}} \frac{\alpha^{N_e}}{\sqrt{N_e!}}$.
During the sweeping, the time-evolution of the system state $\Psi(t)$ obeys the Schr\"{o}dinger equation,
\begin{equation}\label{SE}
    i \frac{\partial |\Psi(t)\rangle}{\partial t}=H(t)|\Psi(t)\rangle,
\end{equation}
where $H(t)$ is the Hamiltonian~\eqref{H2} with time-dependent detuning $\delta(t)$.
The system state can be expanded as
\begin{equation}\label{Psi_ini}
|\Psi(t)\rangle=\sum_{N_e=0}^{\infty}w_{N_e}\left(\sum_{l}C_{l}^{N_e}(t)|N_e-l,-\frac{N_a}{2}+l\rangle\right),
\end{equation}
where $C_{l}^{N_e}(t)$ is the time-dependent coefficient of the basis in subspace with $N_e$ excitation number.
There are $N_e$ subspaces with different excitation numbers and the bases of each subspace depend on the excitation numbers.
Each subspace with different $N_e$ is orthogonal and decoupled with others. Therefore,
the dynamics in each subspace is independent and we can deal with the time-evolution problem individually and then sum up together according to the weight factor of each subspace $w_{N_e}$.
In addition, the average excitation number of the system is always conserved and equals the initial average photon number.

In our calculation, we fix the atom number $N_a=4$, coupling strength $g=1$ and atom-atom interaction $E_c=100$.
Meanwhile, the detuning $\delta$ is linearly swept according to $\delta(t)=\delta_0+\beta t$ with the initial detuning $\delta_0<0$ and the sweeping rate $\beta=\frac{2|\delta_0|}{\tau}$, where $\tau$ is the total sweeping time.
Here, we use $\beta$ to characterize the non-adiabaticity of the sweeping. The smaller sweeping rate $\beta$ refers to a slower sweeping of detuning, which means more probability staying in the instantaneous ground state when the LZ process occurs.
The properties of LZ dynamics with multiple excitation numbers may differ according to the distribution of the excitation numbers which is related to the average photon number of the initial coherent state.
Here, for initial coherent states, we choose the average photon number to be $|\alpha|^2=4$ and $|\alpha|^2=25$.

\subsection{Asymmetric sequential Landau-Zener dynamics}

The evolution of the photon number, the variance of relative atom number and the half relative atom number versus the detuning $\delta(t)$ for different sweeping rates $\beta$ are shown in Fig.~\ref{Fig-Dynamical-LZ}.
The three observables of the evolved state can be respectively calculated according to
\begin{equation}\label{PhoNum}
\langle n(t)\rangle=\langle \Psi(t)|\hat a^{\dagger} \hat a|\Psi(t)\rangle=\sum_{N_e,l} (N_e-l) |w_{N_e} C^{N_e}_{l}(t)|^2,
\end{equation}
\begin{eqnarray}\label{ExANum}
\frac{\langle N_{\uparrow}(t)-N_{\downarrow}(t)\rangle}{2}&=&\langle \Psi(t)|\hat J_{z}|\Psi(t)\rangle \nonumber\\
&=&\sum_{N_e,l} (-\frac{N_a}{2}+l) |w_{N_e} C^{N_e}_{l}(t)|^2,
\end{eqnarray}
and
\begin{eqnarray}\label{Variance}
&&\text{Var}(N_{\uparrow}(t)-N_{\downarrow}(t)) \nonumber\\
&=&4\left[\langle \Psi(t)|\hat J_{z}^2|\Psi(t)\rangle-\left(\langle \Psi(t)|\hat J_{z}|\Psi(t)\rangle\right)^2\right] \nonumber\\
&=&4\{\sum_{N_e,l}(-\frac{N_a}{2}+l)^2 |w_{N_e}C^{N_e}_{l}(t)|^2 \nonumber\\
&-&[\sum_{N_e,l}(-\frac{N_a}{2}+l) |w_{N_e} C^{N_e}_{l}(t)|^2]^2\}.
\end{eqnarray}

For $|\alpha|^2=4$, where the average photon number equals the atom number, the population ladder for half relative atom number coincides with the one in adiabatic limit when the sweeping rate is sufficiently small (e.g., $\beta=2$), see Fig.~\ref{Fig-Dynamical-LZ}~(a).
When the sweeping becomes fast (e.g., $\beta=10$), the population ladder oscillates and the heights of the steps become less obvious.
The sequential LZ transitions gradually disappear when the non-adiabatic effect is strong enough for extremely fast driving (e.g., $\beta=100$).
For $|\alpha|^2=25$, where the average photon number is much greater than the atom number, the sequential LZ dynamics versus the non-adiabatic effect is similar, see Fig.~\ref{Fig-Dynamical-LZ}~(b).
However, it does not require such slow driving to attain its adiabatic limit (e.g., $\beta=20$ is slow enough while for $|\alpha|^2=4$ is not).
That is because when the average photon number increases, the Rabi frequency as well as the gaps of the avoided energy level crossings become larger. The analytical analysis for adiabaticity condition is given in the next subsection.
%

The variances of the relative atom number for small and large $|\alpha|^2$ are very different, see Fig.~\ref{Fig-Dynamical-LZ}~(c) and (d).
For the adiabatic sweeping with $|\alpha|^2=25$, the variance $\text{Var}(N_{\uparrow}-N_{\downarrow})$ is similar to the one with large single excitation number shown in Fig.~\ref{Fig-single-excitation}~(f).
While for the adiabatic sweeping with $|\alpha|^2=4$, the variance $\text{Var}(N_{\uparrow}-N_{\downarrow})$ appears two peaks for the first and second LZ transitions and then shows a ladder-like structure with unequal step heights.
If the sequential LZ transitions for small $N_e$ are incomplete, the atomic state is a superposition state of different $m=(N_{\uparrow}-N_{\downarrow})/2$ and the corresponding variance $\text{Var}(N_{\uparrow}-N_{\downarrow})$ becomes large.
Thus, as the detuning increases, the systems is dominated by the incomplete sequential LZ transitions for small $N_e$, and the atomic number variance changes from peaks to ladder-like steps.
While for large $|\alpha|^2$, the atomic state in every subspace changes synchronously and the variance would appear a series of peaks due to the quasi-degeneracy.

When taking the non-adiabatic effects into account, the variances change dramatically.
For both $|\alpha|^2=4$ and $|\alpha|^2=25$, under a very slow sweeping, the variances approach to their
adiabatic limits.
Under intermediate sweeping, the atomic state becomes a superposition state of different
$m=(N_{\uparrow}-N_{\downarrow})/2$ after every LZ transitions.
As the LZ transitions occur sequentially, more different $m=(N_{\uparrow}-N_{\downarrow})/2$ components accumulate and induce the variance $\text{Var}(N_{\uparrow}-N_{\downarrow})$ to increase in ladder-like steps (e.g., the blue dash-dotted lines in Fig.~\ref{Fig-Dynamical-LZ}~(c) and (d)).
While for the fast sweeping, the system states hardly follow the instantaneous ground state and change slightly and the variances change more smoothly.

More interestingly, the sequential population ladders for small average photon number exhibit another kind of asymmetric property. The heights (but not the slopes) of the steps are obviously unequal and become fractional, see Fig.~\ref{Fig-Dynamical-LZ}~(a) and (e).
For $|\alpha|^2=4$, even though in the adiabatic limit, the height of each step is less than one and different with others. The relative atom population ladder increases upstairs with the height of the steps gradually becomes smaller and smaller.
The 4 photons on average could not be entirely absorbed by the 4 atoms even after 4 times of LZ transitions.

In contrast, for $|\alpha|^2=25$ in adiabatic limit, the population ladder of half relative atom number increases
upstairs with the each step's height equals nearly one, which is similar to the LZ process with single excitation numbers shown in Fig.~\ref{Fig-single-excitation}.
The coherent state with large average photon number can be approximately treated as a classical field.
When the average photon number is large enough, the initial atoms in the lower level could gradually absorb the photon one by one through every LZ transitions, , see Fig.~\ref{Fig-Dynamical-LZ}~(b) and (f).

It is worth to mention that, when $\delta(t)$ is swept adiabatically to 0, the evolved atomic state becomes a twin Fock state, in which the populations in upper and lower levels are exactly equal ($\langle N_{\uparrow}-N_{\downarrow}\rangle=0$) and the variance of the relative atom number $\text{Var}(N_{\uparrow}-N_{\downarrow})=0$.
The atomic twin Fock state is a highly entangled state that can be used to implementing quantum metrology~\cite{Lucke2011, Gerry2010}.
In our system, preparing the initial atomic state into $|-\frac{N_a}{2}\rangle$ and the cavity photon field into a coherent state with large $|\alpha|^2$, an atomic twin Fock state can be generated via an adiabatic linear sweeping from negative detuning to zero.
This offers a new scheme for quantum state engineering on highly entangled state preparation.

\subsection{Analytical analysis for sequential Landau-Zener dynamics}
\begin{figure}[htb]
\centering
\includegraphics[width=\columnwidth]{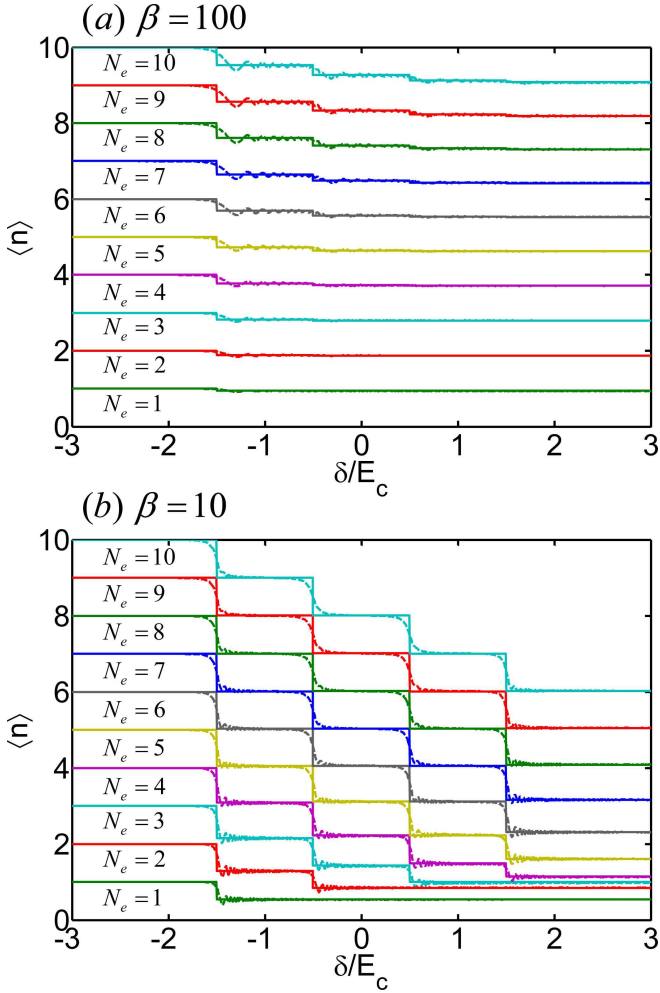}\caption{(Color online) The sequential LZ dynamics for photon number
 in subspaces under different sweeping rate $\beta$. Here, $N_e$ are chosen from $1$ to $10$, $N_a$=4, $g=1$ and
 $E_c=100$. The solid lines are obtained analytically according to Eq.~\eqref{Photon_LZ}. The dashed lines are
 the numerical results.}
\label{Fig-LZ}
\end{figure}

To explore the sequential LZ dynamics, we derive an analytical formula for ladders of the photon number and the relative atom number.
In our model, the whole Hilbert space can be divided into several decoupled subspaces and we can investigate the dynamics individually in every subspace and finally sum up together.
For a subspace with excitation number $N_e$ and atom number $N_a$, it would respectively occur $N_a$ and $N_e$ times of LZ transitions for $N_e\ge N_a$ and $N_e< N_a$.
The $k$-th LZ transition happens between the states $|N_e-k+1, -\frac{N_a}{2}+k-1\rangle$ and $|N_e-k, -\frac{N_a}{2}+k\rangle$ with $k=1,...,\min(N_a,N_e)$.
Therefore, the sequential LZ transitions in a subspace can be treated as a sequence of conventional two-level LZ transitions.
In a conventional LZ transition of a sweeping rate $\beta$ with the minimum gap $\Delta$ for its avoided energy level crossing, and starting from its ground state at time $t\rightarrow-\infty$, the probability of finding the system remaining in the ground state at time $t\rightarrow+\infty$ is given by the LZ formula
~\cite{Landau1932, Zener1932},
\begin{equation}\label{LZ_formula}
P(\Delta,\beta)=1-\exp(-\frac{\pi\Delta^2}{2\beta}).
\end{equation}

By applying the conventional two-level LZ formula to each avoided energy level crossing, the probability of staying in the instant ground state at $k$-th LZ transition reads
\begin{equation}\label{P_LZ}
P(\Delta_k^{\!N_e},\beta)=1-\exp\left[-\frac{\pi\left(\Delta_k^{\!N_e}\right)^2}{2\beta}\right].
\end{equation}
where $\Delta_k$ denotes the minimum gap of the $k$-th LZ transitions, which corresponds to the energy gap at $\delta_k/E_c=-(N_a+1)/2+k$ with $k=1,2,...,\min(N_a,N_e)$.
The $k$-th minimum gap is proportional to the atom-photon coupling term of the Hamiltonian~\eqref{H3}, and it can be expressed in the form of
\begin{equation}\label{min_gap}
  \Delta_k^{\!N_e}=2g\sqrt{\frac{k(N_a-k+1)(N_e-k+1)}{N_a}}.
\end{equation}

By applying the LZ formula~\eqref{P_LZ} one by one, we can obtain the final probability of staying in
the instant ground state after $k$ times of LZ transitions, which is given as
\begin{equation}\label{final_pro}
  P_{N_e,k}(\beta)=\prod_{l=1}^{k} P(\Delta_l^{\!N_e},\beta).
\end{equation}
For every LZ transition, the probability of remaining in the ground state also corresponds to the probability of absorbing one photon.
From this point, we can figure out the heights of every population steps of the photon number during the sweeping.
The height of the $k$-th step equals exactly $P_{N_e,k}(\beta)$.
Therefore, the photon number of the plateaus in the ladder can be obtained analytically,
\begin{equation}\label{Photon_LZ}
  \langle n \rangle_{\!N_e}=\left\{\begin{array}{lll}
    N_e, \delta<\delta_1 \\
    N_e-\sum_{m=1}^{\min(\!k,N_e\!)}P_{N_e,m}(\beta), \delta_k\le\delta<\delta_{k+1}\\
    N_e-\sum_{m=1}^{\min(\!N_e,N_a\!)}P_{N_e,m}(\beta), \delta\ge\delta_{N_a}\\
  \end{array}\right.
\end{equation}
Correspondingly, the half relative atom number can be expressed as
\begin{multline}\label{Jz_LZ}
\frac{\langle N_{\uparrow}-N_{\downarrow}\rangle_{\!N_e}}{2}\\
=\left\{\begin{array}{lll}
    -\frac{N_a}{2}, \delta<\delta_1 \\
    -\frac{N_a}{2}+\sum_{m=0}^{\min(\!k,N_e\!)}P_{N_e,m}(\beta), \delta_k\le\delta<\delta_{k+1}\\
    -\frac{N_a}{2}+\sum_{m=0}^{\min(\!N_e,N_a\!)}P_{N_e,m}(\beta), \delta\ge\delta_{N_a}\\
  \end{array}\right.
\end{multline}
where $\delta_k=\left[-(N_a+1)/2+k\right]E_c$ and the index $k=1,2,...,\min(N_a,N_e)$.
Given a sweeping rate $\beta$, one can figure out the population ladders analytically. In Fig.~\ref{Fig-LZ}, we plot the population ladders of photon number for $N_e=\{1, 2, ..., 10\}$ under sweeping rates $\beta=\{10,100\}$.
The dashed lines are the numerical results while the solid lines are obtained analytically according to Eq.~\eqref{Photon_LZ}. It is obvious that, the analytical population ladders are well consistent with the numerical ones.
Since the minimum gap $\Delta_{k}$ increases with excitation number $N_e$, it needs slower driving for smaller $N_e$. For fast sweeping $\beta=100$, the step height decreases obviously when $N_e$ changes from $10$ to $1$. While for slow sweeping $\beta=10$, the step heights for most $N_e$ approach to 1.
Therefore, for small $|\alpha|^2$, the dominated $N_e$ is small and requires slower sweeping to access the adiabaticity condition.

By summing up the results of all subspaces, the average photon number and the half relative atom number of the system are given as
\begin{equation}\label{sum_ph}
    \langle n \rangle= \sum_{N_e=0}^{\infty} w_{N_e}^2 \langle n \rangle_{N_e},
\end{equation}
and
\begin{equation}\label{sum_at}
    \frac{\langle N_{\uparrow}-N_{\downarrow}\rangle}{2}= \sum_{N_e=0}^{\infty} w_{N_e}^2 \frac{\langle N_{\uparrow}-N_{\downarrow}\rangle_{\!N_e}}{2}.
\end{equation}
In our system, the adiabatic condition can be estimated approximately according to
\begin{equation}\label{AdCondition}
\frac{\left[\min_k\left(\Delta_{k}^{|\alpha|^2-\min(\!|\alpha|^2,N_a\!)+1}\right)\right]^2}{2\beta} \gtrsim 2.
\end{equation}
For $|\alpha|^2=4$, $\beta\lesssim1$; for $|\alpha|^2=25$, $\beta\lesssim21$. This estimation can be confirmed by the comparison with numerical calculations, see Fig.~\ref{Fig-Dynamical-LZ}.

One of the interesting findings is the appearance of fractional steps in population ladders for small $|\alpha|^2$
even under adiabatic sweeping.
This peculiar phenomenon results from the property of coherent state with few average photon number. The initial coherent state is a superposition of infinite Fock state including the vacuum state with different weight coefficients $w_{N_e}$.
For small $N_e$, there is no sufficient photons to excite all the atoms to the upper level, and this incomplete sequential LZ process would contribute to the fractional steps if the weight coefficient $w_{N_e}$
is non-ignorable.

Below we give an analytical explanation for the fractional steps.
When the sweeping is adiabatic ($\beta\rightarrow 0$), all $P_{N_e,k}(\beta)\rightarrow 1$ and so that the average photon number in subspace of $N_e$ becomes
\begin{equation}\label{ph_ad}
  \langle n \rangle_{\!N_e}=
  \left\{\begin{array}{lll}
    N_e, \delta<\delta_1 \\
    N_e-\text{min}(k,N_e), \delta_k\le\delta<\delta_{k+1}\\
    N_e-\text{min}(N_e,N_a), \delta\ge\delta_{N_a}\\
  \end{array}\right.
\end{equation}
According to Eq.~\eqref{sum_ph}, the average photon number of the system can be obtained,
\begin{equation}\label{ph_coh_ad}
  \langle n \rangle=
  \left\{\begin{array}{lll}
    \sum_{N_e=0}^{\infty} w_{N_e}^2 N_e, \delta<\delta_1 \\
    \sum_{N_e=k}^{\infty} w_{N_e}^2(N_e-k), \delta_k\le\delta<\delta_{k+1}\\
    \sum_{N_e=N_a}^{\infty} w_{N_e}^2(N_e-N_a), \delta\ge\delta_{N_a}\\
  \end{array}\right.
\end{equation}
Using the condition that $\sum_{N_e=0}^{\infty} w_{N_e}^2 N_e=|\alpha|^2$, we can simplify Eq.~\eqref{ph_coh_ad} to
\begin{equation}\label{ph_coh_ad1}
  \langle n \rangle=
  \left\{\begin{array}{lll}
    |\alpha|^2, \delta<\delta_1 \\
    |\alpha|^2-k-\sum_{N_e=0}^{k-1} w_{N_e}^2(N_e-k), \delta_k\le\delta<\delta_{k+1}\\
    |\alpha|^2-N_a-\sum_{N_e=0}^{N_a-1}w_{N_e}^2(N_e-N_a), \delta\ge\delta_{N_a}\\
  \end{array}\right.
\end{equation}
From Eq.~\eqref{ph_coh_ad1}, we can further obtain the $k$-th step height of the photon number (as well as the relative atom number),
\begin{equation}\label{stepH}
h_{k}=1-\sum_{N_e=0}^{k-1}w_{N_e}^2.
\end{equation}
As a result, for small $|\alpha|^2$, the weight coefficients $w_{N_e}^2$ of small $N_e$ are dominated, which leads to the fractional steps.
For example, $|\alpha|^2=4$, $w_{0}^2=0.0183$, $w_{1}^2=0.0733$, $w_{2}^2=0.1465$, and $w_{3}^2=0.1954$ and we can get $h_1=0.9817$, $h_2=0.9084$, $h_3=0.7619$ and $h_4=0.5665$. The results are the same with the numerical findings.
While for large $|\alpha|^2$,  the weight coefficients $w_{N_e}^2$ of small $N_e$ are nearly zero, and the height of every step equals 1.
Since the total excitation number of the system is conserved, the explanation is also valid to the relative atom number.


\subsection{Dissipative Sequential Landau-Zener dynamics}
\begin{figure*}[htb]
\centering
\includegraphics[scale=0.6]{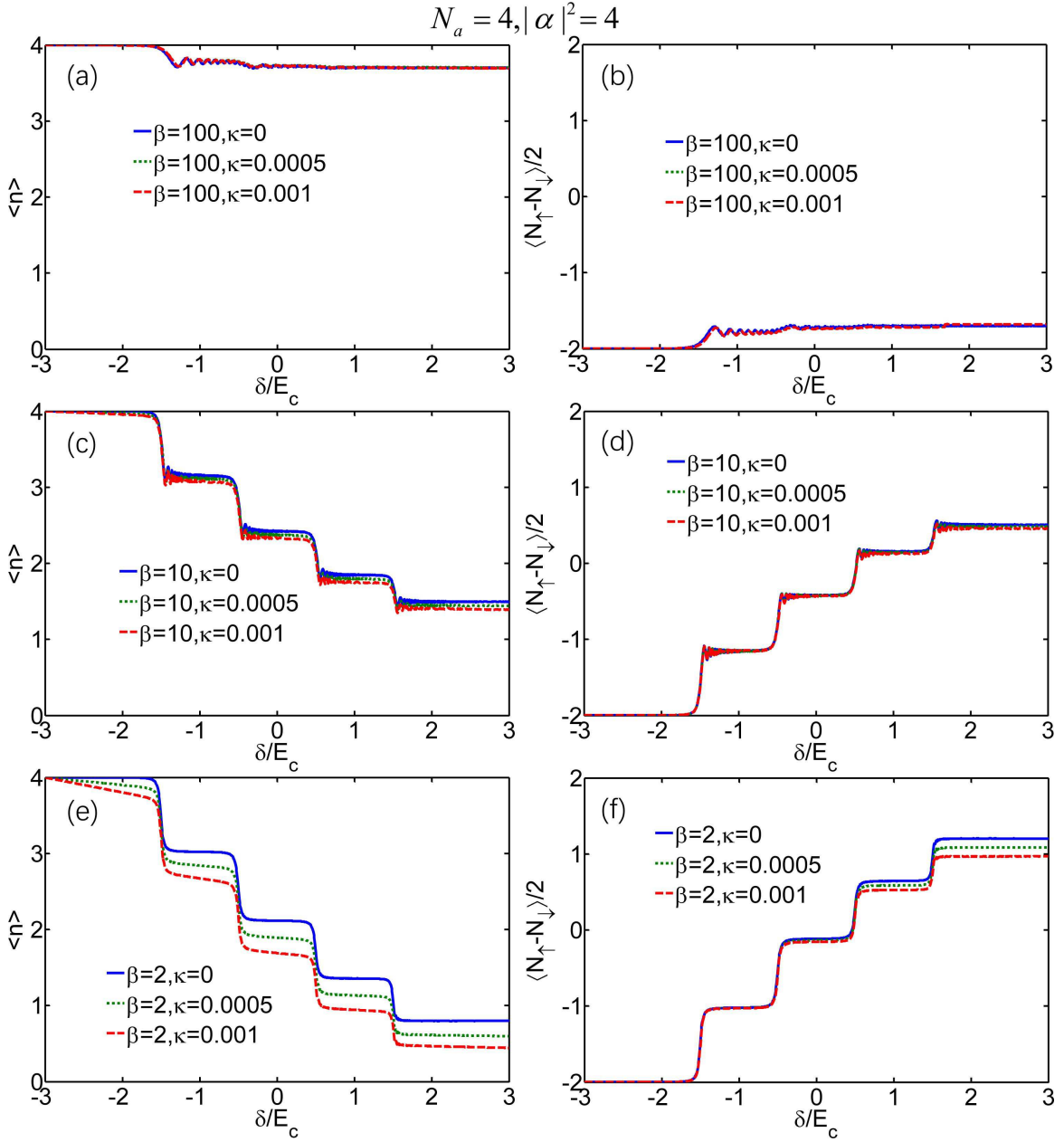}\caption{(Color online) The sequential LZ dynamics under cavity phton losses. The influences of damping rates $\kappa$ on the photon number and the relative atom number under different sweeping rates $\beta$ are shown. Here, the parameters are chosen as $N_a=4$, $|\alpha^2|=4$, $g=1$, and $E_c=100$.}
\label{Fig-dissipation}
\end{figure*}

In cavity QED experiments, the system would suffer decoherence induced by the escape of photons out of the cavity or the decay of the atoms without emitting photons~\cite{Reiserer2015, Hartmann2008}.
These decoherence effects may affect the sequential LZ transitions.
In this subsection, we mainly discuss the effect of dissipation (cavity photon loss) on the sequential LZ dynamics.
During the sweeping, the cavity photons may escape from the cavity and the total excitation number would no longer be conserved.
Therefore, it is necessary to describe the system state by a reduced density matrix $\rho=|\tilde{\Psi}\rangle\langle\tilde{\Psi}|$, where the bases of $|\tilde{\Psi}\rangle$ include different total excitation numbers and differ from the ones in Eq.~\eqref{Psi_ini}. The sequential dynamics under cavity photon loss can be characterized by the Lindblad master equation~\cite{Keeling2008, Hartmann2008, Reiserer2015}
\begin{equation}\label{MEq}
\frac{\partial \rho}{\partial t}=-i\left[ H(t), \rho \right]+L_{\kappa}[\rho],
\end{equation}
with
\begin{equation}\label{Kappa}
L_{\kappa}[\rho]=\frac{\kappa}{2}\left( 2\hat a \rho \hat a^{\dagger} - \hat a^{\dagger} \hat a \rho -\rho \hat a^{\dagger} \hat a \right),
\end{equation}
where $\kappa$ is the cavity photon loss rate.

In our calculation, the reduced density matrix of the system is broken up into different density matrices $\rho_{N_e}$
in subspaces.
Then we solve the master equation~\eqref{MEq} independently and calculate the observable expectation by $\langle \hat O_{\!N_e}\rangle=Tr\left(\hat O \rho_{N_e}\right)$ and sum up together according to $\langle \hat O\rangle=w_{N_e}^2 \langle \hat O_{\!N_e}\rangle$. The time-evolution of photon number and half relative atom number under dissipation for $|\alpha|^2=4$ are shown in Fig.~\ref{Fig-dissipation}.

For $\kappa=0$, it returns to the ideal case, where the total excitation number is conserved.
For $\kappa>0$, the influences of the dissipation become obvious when the sweeping rates getting smaller.
Given $\beta=100$ and $\beta=10$, the influences of dissipation are very small and the sequential LZ dynamics
are almost unchanged, see Fig.~\ref{Fig-dissipation}~(a)-(d).

When the sweeping becomes slower, the total evolution time becomes longer and more photon losses accumulate.
For $\beta=2$, which approaches to the adiabatic limit for $|\alpha|^2=4$, although the cavity photon loss decreases the photon number, it only changes the step heights of the fractional ladders slightly, see Fig.~\ref{Fig-dissipation}~(e).
The cavity photon loss affects the ladder of the half relative atom number gradually, but the step heights shrink only a little during the whole processes, see Fig.~\ref{Fig-dissipation}~(f).

It is shown that, the cavity photon loss may also affect the sequential population ladders.
Obviously, the sequential population ladders will disappear if the lost photons is large compared with $|\alpha|^2$.
However, if the lost photons is small compared with $|\alpha|^2$, which may be realized in the strong coupling regime~\cite{Reiserer2015, Hartmann2008}, the fractional steps of half relative atom number can still be observed.

\subsection{Collapse and revivals of cavity field coherence}
\begin{figure*}[htb]
\centering
\includegraphics[scale=0.7]{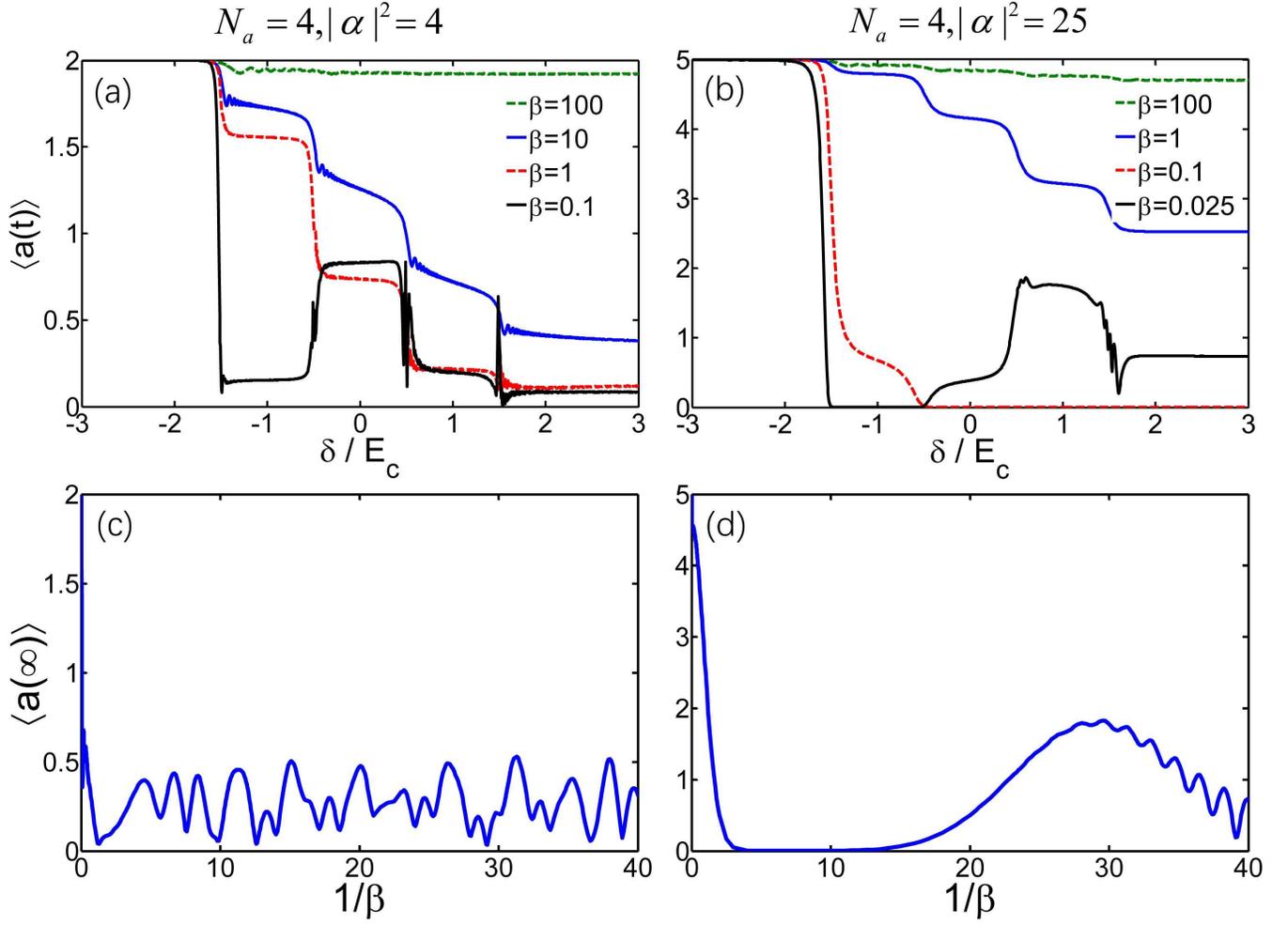}\caption{(Color online) Collapse and Revivals of cavity field coherence. The evolution of cavity field coherence $\langle a(t) \rangle$ during LZ dynamics with different sweeping rate $\beta$ for (a) $|\alpha|^2=4$ and (b) $|\alpha|^2=25$. The final cavity field coherence $\langle a(\infty)\rangle$ versus the inverse of sweeping rate $1/\beta$ for (c) $|\alpha|^2=4$ and (d) $|\alpha|^2=25$. Here, $N_a=4$, $g=1$ and $E_c=100$.}
\label{Fig-Collapse-revivals}
\end{figure*}

We further investigate how the coherence of the cavity field varies during the LZ processes with different sweeping rates.
Since the evolution of each photon Fock state will accumulate a phase dependent on the photon number, the phase differences between different Fock state will also play an important role during the LZ dynamics.
In order to study the effect of this phase difference during the evolution, a common quantity to measure is the expectation of the photon filed annihilation operator~\cite{Keeling2008}
\begin{eqnarray}\label{Coh}
\langle a(t)\rangle&=&\langle \Psi(t)|\hat a|\Psi(t)\rangle \nonumber\\
&=&\sum_{N_e,l} \sqrt{N_e-l}w^{*}_{N_e-1}\! w_{N_e} C^{N_e-1*}_{l}\!(t) C^{N_e}_{l}\!(t),
\end{eqnarray}
which well describes the coherence of the cavity field.
The coherence of the photon filed $\langle a(t)\rangle$ versus the evolution time $t$ with different sweeping rates $\beta$ are shown in Fig.~\ref{Fig-Collapse-revivals}~(a) and (b).
The dynamical behaviors of $\langle a(t)\rangle$ are much different with the ones of photon number $\langle n(t) \rangle$.

When $|\alpha|^2=25$, for very fast sweeping $\beta=100$, the coherence of the cavity field $\langle a(t)\rangle$ changes slightly.
For modest sweeping $\beta=1$, $\langle a(t)\rangle$ oscillates in ladder shape.
For slower sweeping $\beta=0.1$, $\langle a(t)\rangle$ drops dramatically and remains zero when passing through all the LZ transitions.
However, for very slow sweeping $\beta=0.025$, $\langle a(t)\rangle$ drops rapidly to zero when the first LZ transition happens and begins to increase and oscillate after the second LZ transition. Finally, the coherence remains a certain value after all the LZ transitions.
This is some kind of collapse and revivals of the cavity field coherence due to the long time accumulated phase difference between different Fock state components of the coherent state.

The changes of the cavity field coherence for $|\alpha|^2=4$ is similar with the one of $|\alpha|^2=25$.
For fast sweeping, the coherence of the photon filed $\langle a(t)\rangle$ changes slightly with small oscillations.
For modest sweeping, $\langle a(t)\rangle$ oscillates in a downstairs ladder-like shape.
For very slow sweeping, the coherence of the cavity field $\langle a(t)\rangle$ first collapses and then revives. It drops quickly at the first LZ transition point and oscillates a lot afterwards until all the LZ transitions finished.
However, for $|\alpha|^2=4$, the coherence of the cavity field seldom decreases to zero and the oscillations are more dramatic compared with the ones of $|\alpha|^2=25$.

The coherence of the cavity field will tend to a steady value after all the LZ transitions and the final coherence $\langle a(\infty)\rangle$ depends on the sweeping rate.
Below the dependence of $\langle a(\infty)\rangle$ on the inverse sweeping rate $1/\beta$ is shown in Fig.~\ref{Fig-Collapse-revivals}~(c) and (d).
Obviously, the final coherence of the cavity field versus the inverse sweeping rate also exhibits the behavior of collapse and revivals.
For $|\alpha|^2=25$, with fast sweeping, the collapse occurs as the difference between different Fock state grows leading to a destructively interference. With slower sweeping, the coherence drops to death and remains zero for a wide range. For much slower sweeping, the phase differences grows back in phase and the coherence gradually begins to revive. However, the revival will oscillate and finally cease for extremely slow sweeping.
While for $|\alpha|^2=4$, the coherence changes more rapidly. The final coherence of the cavity field drops dramatically at first but revives quickly and then begins to oscillate in disorder around a modest range. It may be seen that, the cavity field coherence for small average photon number would not exhibit obvious collapse and revivals, which is different from the one for large average photon number.

\section{Summary and Discussions}\label{Sec5}

In summary, we have explored the novel phenomena of asymmetric sequential LZ dynamics in an ensemble of  interacting two-level Bose condensed atoms trapped in an optical cavity.
The features of Bose condensed atoms in a cavity with small photon number are extremely different from the ones with large photon number.
For relatively small photon number, the interplay between the detuning and the atom-atom interaction leads to asymmetric sequential LZ transitions.
There appear asymmetric interaction blockade and single-atom resonance tunneling when the LZ process involves only a single excitation number, which is mainly due to the photon-number-dependent Rabi frequency.
The single-atom resonance may be used to design high-precision devices and sensitive detectors.

Instead of an initial Fock state, considering an initial coherent cavity field with small average photon number, an asymmetric population ladder with unequal fractional height steps is observed.
The asymmetric sequential LZ dynamics of the system are also studied.
We derive an analytical formula for sequential population ladders. The adiabaticity conditions for the sequential LZ transitions are analytically estimated. An intriguing findings of fractional steps are also explained. In addition, we analyze the effects of the cavity dissipation on the sequential LZ dynamics.
On the other hand, the state of the cavity field changes dramatically during the time-evolution process.
The behaviors of collapse and revivals of the cavity field coherence are revealed.

Further, for an initial coherent state with large average cavity-photon number, the sequential LZ dynamics in a cavity approach to the one in a continuous laser field.
Starting from all atoms in the lower energy level, the atoms can absorb photons one by one in the sequential LZ process if the detuning is swept sufficiently slow.
When the detuning is swept to zero, an atomic twin Fock state can be generated.
In addition, the photon state is no longer a coherent state and its coherence changes dramatically.
This may be associated to the generation of photon-deleted coherent states~\cite{Honer2011}, which is totally non-classical.
Similarly, if one sweeps the detuning oppositely, and the atoms are initially in the upper level, there would be the phenomenon that the atoms emit photons one by one during the sequential LZ process.
This may be related to the creation of the photon-added coherent states~\cite{Noriyuki2011}.
Different from most previous schemes which add or subtract only a single photon, our this scheme can add or subtract desired number of photons by dynamically controlling the detuning.
This indicates that it may provide a new tool for preparing non-classical photon states and also may be applied to demonstrate the bosonic commutation relations~\cite{Zavatta2004,Parigi2007,Zavatta2008}.

\section*{Acknowledgements}

This work is supported by the National Basic Research Program of China (NBRPC) under Grant No. 2012CB821305, the National Natural Science Foundation of China (NNSFC) under Grants No. 11374375, No. 11465008, and No. 11574405.


\end{document}